\documentclass[useAMS,natbib,multicolumn,subfigure]{mn2e}

\usepackage{amsmath}
\usepackage[T1]{fontenc}
\usepackage{ae,aecompl}

\usepackage[latin1]{inputenc}  % ACETUACAO EM PORTUGUES
\usepackage{graphicx,amsmath,xcolor}
\usepackage{amssymb}
\usepackage{indentfirst}
\usepackage{hyperref}
\usepackage{graphicx,pst-all}
\usepackage{tikz}
\usepackage{layout}
\usepackage{fancyhdr}

\def\msol{\hbox{\kern 0.20em $M_\odot$}}
\def\lsol{\hbox{\kern 0.20em $L_\odot$}}
\def\rsol{\hbox{\kern 0.20em $R_\odot$}}
\def\sr{\hbox{\kern 0.20em sr}}
\def\srmu{\hbox{\kern 0.20em sr$^{-1}$}}

\def\g{\hbox{\kern 0.20em g}}
\def\gmu{\hbox{\kern 0.20em g$^{-1}$}}
\def\kg{\hbox{\kern 0.20em kg}}
\def\pc{\hbox{\kern 0.20em pc}}

\def\mum{\hbox{\kern 0.20em $\mu$m}}
\def\mumd{\hbox{\kern 0.20em $\mu$m$^{-2}$}}
\def\cm{\hbox{\kern 0.20em cm}}
\def\m{\hbox{\kern 0.20em m}}
\def\km{\hbox{\kern 0.20em km}}
\def\nm{\hbox{\kern 0.20em nm}}

\def\s{\hbox{\kern 0.20em s}}
\def\h{\hbox{\kern 0.20em h}}
\def\sec{\hbox{\kern 0.20em sec}}
\def\min{\hbox {\kern 0.20em min}}
\def\smu{\hbox{\kern 0.20em s$^{-1}$}}
\def\smd{\hbox{\kern 0.20em s$^{-2}$}}
\def\an{\hbox{\kern 0.20em an}}
\def\anmu{\hbox{\kern 0.20em an$^{-1}$}}
\def\deg{\hbox{\kern 0.20em $^{\rm o}$}}
\def\yr{\hbox{\kern 0.20em yr}}
\def\yrmu{\hbox{\kern 0.20em yr$^{-1}$}}
\def\Myr{\hbox{\kern 0.20em Myr}}
\def\Mymu{\hbox{\kern 0.20em Myr$^{-1}$}}
\def\K{\hbox{\kern 0.20em K}}
\def\pcmu{\hbox{\kern 0.20em pc$^{-1}$}}
\def\pcmd{\hbox{\kern 0.20em pc$^{-2}$}}
\def\pcmt{\hbox{\kern 0.20em pc$^{-3}$}}
\def\kms{\hbox{\kern 0.20em km\kern 0.20em s$^{-1}$}}
\def\kmpd{\hbox{\kern 0.20em km$^{2}$}}
\def\kpc{\hbox{\kern 0.20em kpc}}
\def\cms{\hbox{\kern 0.20em cm\kern 0.20em s$^{-1}$}}
\def\erg{\hbox{\kern 0.20em erg}}
\def\ergs{\hbox{\kern 0.20em erg}}
\def\cmpd{\hbox{\kern 0.20em cm$^2$}}
\def\cmmd{\hbox{\kern 0.20em cm$^{-2}$}}
\def\cmms{\hbox{\kern 0.20em cm$^{-6}$}}
\def\cmpt{\hbox{\kern 0.20em cm$^3$}}
\def\cmmt{\hbox{\kern 0.20em cm$^{-3}$}}
\def\mpd{\hbox{\kern 0.20em m$^2$}}
\def\mmd{\hbox{\kern 0.20em m$^{-2}$}}
\def\mpt{\hbox{\kern 0.20em m$^3$}}
\def\mmt{\hbox{\kern 0.20em m$^{-3}$}}
\def\mujy{\hbox{\kern 0.20em $\mu$Jy}}
\def\mjy{\hbox{\kern 0.20em mJy}}
\def\Mj{\hbox{\kern 0.20em MJy}}
\def\jy{\hbox{\kern 0.20em Jy}}
\def\ghz{\hbox{\kern 0.20em GHz}}
\def\srmd{\hbox{\kern 0.20em sr$^{-1}$}}

\def \mum{$\mu$m}

\def\G{\hbox{\kern 0.20em G}}

\def\htwo{\hbox{H${}_2$}}
\def\h13cop{\hbox{H$^{13}$CO$^{+}$}}

\def\h2o{\hbox{H$_2$O}}

\title[P-bearing molecules in L1157-B1]{Phosphorus-bearing molecules in solar-type star forming regions: First PO detection.}

\author[B. Lefloch et al.]{
Bertrand Lefloch$^{1,2,3}$\thanks{E-mail: bertrand.lefloch@univ-grenoble-alpes.fr},
C. Vastel$^{4,5}$,
S. Viti$^6$,
I. Jimenez-Serra$^{6,7}$,
C. Codella$^{8}$,
L. Podio$^{8}$,
\newauthor
C. Ceccarelli$^{1,2}$,
E. Mendoza$^{3,1}$,
J.R.D. Lepine$^3$,
R. Bachiller$^9$
\\
$^{1}$Univ. Grenoble Alpes, IPAG, F-38000, Grenoble, France\\
$^{2}$CNRS, IPAG, F-38000 Grenoble, France\\
$^{3}$IAG, Universidade de S\~ao Paulo, Cidade Universit\'aria, SP 05508-090, Brazil \\
$^{4}$Universit\'e de Toulouse, UPS-OMP, IRAP, 31028, Toulouse Cedex 4, France\\
$^{5}$CNRS, IRAP, 9 Av. colonel Roche, BP 44346, 31028, Toulouse Cedex 4, France\\
$^6$Department of Physics and Astronomy, University College London, Gower Street, London, WC1E 6BT, England \\
$^7$School of Physics and Astronomy, Queen Mary, University of London, Mile End Road, London E1 4NS, England\\
$^8$INAF, Osservatorio Astrofisico di Arcetri, Largo Enrico Fermi 5, I-50125 Firenze, Italy\\
$^9$Observatorio Astron\'omico Nacional (OAN), Apdo 112, 28803 Alcal\'a de Henares, Madrid, Spain
}

\date{Accepted 2016 July 28. Received 2016 July 28; in original form 2016 June 13}

\pubyear{2016}

\begin{document}
\label{firstpage}
\pagerange{\pageref{firstpage}--\pageref{lastpage}}
\maketitle

\begin{abstract}
As part of the Large Program ASAI (Astrochemical Surveys At IRAM), we have used the IRAM 30m telescope to lead a systematic search for the emission of rotational transitions of P-bearing species between 80 and 350 GHz towards L1157-B1, a shock position in the solar-type  star forming region L1157. We report the detection of several transitions of PN and, for the first time, of prebiotic molecule PO.  None of these species are detected towards the driving protostar of the outflow L1157-mm.  Analysis of the line profiles shows that PN arises from the outflow cavity, where SiO, a strong shock tracer, is produced.  Radiative transfer analysis yields an abundance  of $2.5\times 10^{-9}$ and $0.9\times 10^{-9}$ for PO and PN, respectively. These results imply a strong depletion ($\approx 100$) of Phosphorus in the quiescent cloud gas. Shock modelling shows that atomic N plays a major role in the chemistry of PO and PN. The relative abundance of PO and PN brings constraints both  on the duration of the pre-shock phase, which has to be $\sim 10^6\yr$, and on the shock parameters. The maximum temperature in the shock has to be larger than $4000\K$, which implies a shock velocity of  $40\kms$.
\end{abstract}


%%%%%%%%%%%%%%%%%%%%%%%%%%%%%%%%%%%%%%%%%%%%%%%%%%

%%%%%%%%%%%%%%%%% BODY OF PAPER %%%%%%%%%%%%%%%%%%

\section{Introduction}

%Recent observational studies on the chemical composition of protostellar shocks have revealed these regions as true astrochemistry %laboratories.
%The presence of Complex Organic Molecules in a protostellar shock region was first reported by  Arce et al. (2008). The molecular content of %tis protostellar shock region (L1157-B1, at d=250\pc, ???), was investigated in the framework  of the IRAM Large Program ASAI dedicated %to chemical surveys of solar-type star-forming regions (Lefloch \& Bachiller in prep.)
%Mendoza et al. (2014) have reported the presence of formamide (NH$_2$CHO), a simple molecule with a peptide bond in shock positions B1 %and B2 in the L1157 outflow. This molecule was proposed by Saladdino et al (2012) to play a key-role in the synthesis of pre-biotic %molecules of genetic and metabolic interest. This result is all the more unexpected as the abundances derived range among the highest ever %reported in the Galaxy. %Formation route is still debated.Interestingly, acetamide, the next most simple amide NH$_2$COCH$_3$, was not %detected towards L1157-B1, with an %upper limit of 0.1 on its abundance reiative to formamide. On the contrary, acetamide was found as %abundant as acetamide towards SgrB2 (Halfen et %al. 2011).
%e

Despite a rather low elemental abundance of $\sim 3\times 10^{-7}$ (Asplund et al. 2009), Phosphorus is one of the main biogenic elements, present in all life forms on Earth. As such, phosphorus-bearing compounds, in particular their P--O bonds,  play a key role in many biochemical and metabolic processes in living systems  (see e.g. Mac\'{\i}a et al. 2005 for a review). In our Solar System, the presence of Phosphorus has been recently reported in comet 67P/Churyumov-Gerasimeno (Altwegg et al. 2016), although the nature of the actual carriers remain to be identified. Phosphorus-bearing compounds appear to be rather ubiquitous in meteorites (Mac\'{\i}a et al. 2005).

A theoretical study by Thorne et al. (1984) based on laboratory experiments  predicted that  Phosphorus monoxyde PO should be the most abundant P-bearing molecule in molecular clouds, hence the main reservoir of phosphorus in the gas phase. Phosphorus monoxyde PO was detected for the first time  by Tenenbaum et al. (2007) towards the evolved star VY Cma. The  P-bearing species HCP, PH$_3$, CP, CCP radicals, PO, and PN have been identified around evolved stars  IRC+10216 (see e.g. Agundez et al. 2007 for a review) and IK Tau (De Beck et al. 2013).  PO was found as abundant as  PN in the envelope of VY CMa, a fact which led the authors to propose that these species are formed from shocks in the circumstellar envelope; they also concluded that PO and  PN  are the main reservoir of Phosphorus in the gas phase.

%Only PN and PO have been identified in molecular clouds.
PN was first detected towards a few high-mass star forming regions: Ori(KL) (Ziurys, 1987),  W51M and SgrB2 (Turner \& Bally 1987). A systematic search for PO by Matthews, Feldman \& Bernath (1987) yielded only upper limits in the massive Star-Forming regions Ori(KL), SgrB2 and DR21(OH).  Recently, Fontani et al. (2016) observed  a sample of 26 massive cores at various evolutionary  stages, from prestellar to ultra-compact HII regions and report detection of the PN $J$=2--1 line in about 30\%  of the their sample. Rivilla et al. (2016) have reported the first detection of PO towards high-mass star forming regions and found that Phosphorus seems strongly depleted from the gas phase.
The first evidence of P-bearing species in low-mass star forming regions was provided by Yamaguchi et al. (2011), who reported the  tentative detection of the PN transition $J$=2--1 towards the shock positions B1 and B2 in the outflow of the low-mass Class 0 protostar  L1157-mm (d=250pc). This is the first and only P-bearing molecule tentatively detected in a solar-type star forming region until now.

Given the importance of Phosphorus for prebiotic chemistry and its presence in the early stages of our own Solar system,  we have led a systematic search for the presence  of phosphorus-bearing molecules in solar-type star forming regions, in the framework of the Large Program dedicated to Astrochemical Surveys At IRAM (ASAI; Lefloch \& Bachiller, in prep), with the 30m telescope. In this Letter, we present the results of  our study towards the outflow shock  L1157-B1 and the driving protostar L1157-mm.  No emission was detected towards the protostar. Towards L1157-B1, we have detected the emission of various rotational transitions of PN and of PO, for the first time.  A search for more complex P-Bearing species (e.g. PH$_3$) yielded only negative results. After presenting the observational results on our systematic search (Section 2), we derive the physical conditions and molecular abundances for both species, and discuss the implication on their formation and shock chemistry.

%% De Beck 2013
%% Thus far, HCP, PH3, the CP and CCP radicals, PO, and PN are the only P-bearing molecules that have been identified around evolved stars.
%% All of these, except PO, were observed towards the CSE of the carbonrich AGB star IRC +10 216 (
%% HCP and PN were also observed in the envelope of the carbon-rich post-AGB object CRL 2688 (Milam et al. 2008), while only PO and PN
%% have thus far been shown to be present in the CSE of an oxygenrich evolved star, namely the red supergiant (RSG) VY CMa
%%
%% Milam et al. (2008)
%%Phosphorus as an element is not particularly abundant, as mentioned, and it is thought to be refractory, as
%% suggested by condensation models (Lodders & Fegley 1999)
%%
%% Assuming solar abundance (P/H  3 ; 107), and neglecting the possible presence of other phosphorus-bearing species, about
%% 7% of the available phosphorus is in the form of gas-phase molecules toward IRC +10216, and about 33% in CRL 2688. In VY CMa, the percentage is %% 21%. The higher amounts of gasp-phase phosphorus in CRL 2688 and VY CMa may result from the energetic outflows present in both of these
%% sources %% (Humphreys et al. 2007; Skinner et al. 1997), which could be destroying phosphorus-bearing grains. Both PN and PO, however, appear %%to arise
%%from the more quiescent wind in VY CMa
%%

\section{Observations}
The observations of L1157-B1 and L1157-mm were acquired with the IRAM-30m telescope at Pico Veleta (Spain),
during several runs in 2011 and 2012. The observed position of L1157-B1 and L1157-mm are  $\alpha_{J 2000} =$ 20$^{\text h}$ 39$^{\text m}$ 10.$^{\text s}$2, $\delta_{J 2000} =$ +68$^{\circ}$ 01$^{\prime}$ 10$^{\prime\prime}$ and $\alpha_{J 2000} =$ 20$^{\text h}$ 39$^{\text m}$ 06.$^{\text s}$3, $\delta_{J 2000} =$ +68$^{\circ}$ 02$^{\prime}$ 15$^{\prime\prime}$.8, respectively.
The survey was carried out  using  the broad-band EMIR receivers at 3~mm (80 -- 116 GHz), 2~mm (128 -- 173 GHz), 1.3~mm (200 -- 272 GHz). Fast Fourier Transform Spectrometers were connected to the EMIR receivers, providing a spectral resolution of 195 kHz. The high-frequency part of the 1.3mm band (260--272 GHz) was observed with the WILMA autocorrelator, at 2 MHz resolution. The final kinematic resolution of the FTS data was degraded to $1\kms$. The observations were carried out in Wobbler Switching Mode, with a throw of $3^{\prime}$, in order to ensure a flat baseline across the spectral bandwith observed (4 GHz to 8 GHz, depending on the receiver).

The data reduction was performed using the GILDAS/CLASS90 package\footnote{http://www.iram.fr/IRAMFR/GILDAS/}. The line intensities are expressed in units of antenna temperature corrected for atmospheric attenuation and rearward losses ($T_A^{\star}$). For the ASAI data, the calibration uncertainties are
typically 10, 15, and $20\%$ at 3mm, 2mm, and 1.3mm, respectively. For subsequent analysis, fluxes were expressed in main beam temperature units ($T_{mb}$). The telescope and receiver parameters (main-beam efficiency Beff, forward efficiency Feff, Half Power beam Width HPBW) were taken from the IRAM webpage\footnote{http://www.iram.es/IRAMES/mainWiki/Iram30mEfficiencies}.
We used in the following the CASSIS software\footnote{http://cassis.irap.omp.eu} for the line identification.

We have summarized the spectroscopic properties and observational parameters of the lines detected in the Table~1. We show in Fig.~1 a montage of the PO and PN lines detected towards L1157-B1.

\section{Results}

\begin{figure}
\begin{center}
\includegraphics[width=1.0\columnwidth]{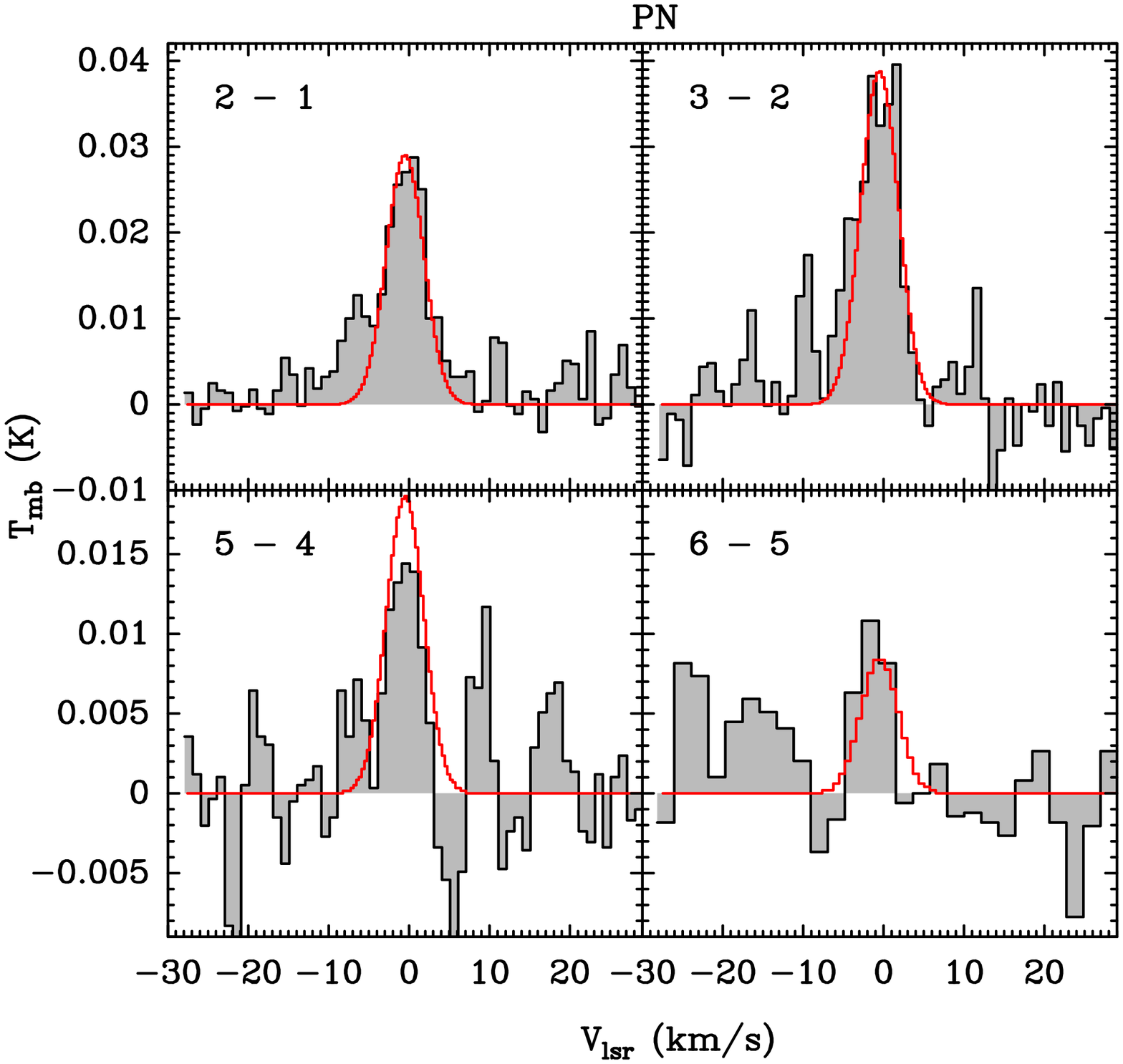}
\includegraphics[width=1.0\columnwidth]{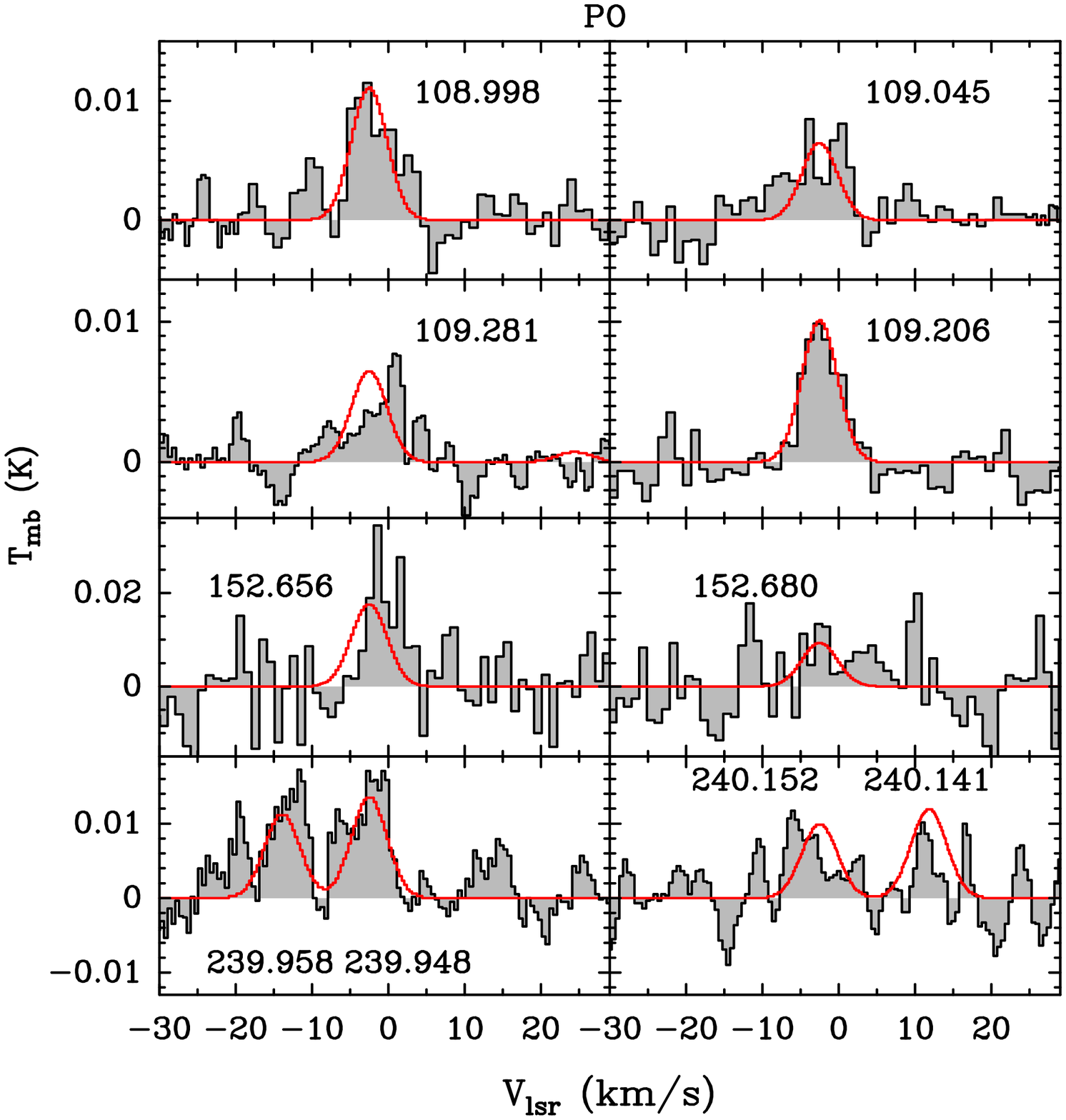}
\caption[]{Montage of the PN and PO lines detected towards protostellar shocks L1157-B1. The red curve represents the best fit obtained from the PO and PN line radiative transfer analysis (see Sect.~4). }
\end{center}
\end{figure}

\subsection{PN}
We used the spectroscopic parameters provided by CDMS for the phosphorus nitride molecule (Cazzoli et al. 2006).
The four PN rotational transitions   $J$=2--1, 3--2, 5--4 and 6--5 fall into the ASAI bands (see Table~1).
No emission at all was detected towards L1157-mm (see the  $J$=2--1 spectrum in red in the bottom panel of Fig.~2). On the contrary,  all of them  were detected towards L1157-B1, with a SNR between 4 and 10. We confirm the previous detection by Yamaguchi et al.  (2011) of the $J$=2--1 and report the detection of the transitions $J$=3--2, 5--4 and 6--5, from levels with upper energy levels $E_{\rm up}$ up to $47\K$. The spectra display a linewidth in the range 4.5--$6.2\kms$, similar to the values measured for several other molecular species (see Codella et al. 2010), which testify of the shock origin of the emission.

The $J$=2--1 and 3--2 lines are detected with a typical intensity of 30~mK. The SNR in the wings of the $J$=2--1 and $J$=3--2 line profiles is high enough to constrain the slope of the intensity distribution as a function of velocity, following the approach of Lefloch et al. (2012) (see also G\'omez-Ruiz et al. 2015). The intensity profile is well described by an exponential law $\propto \exp(v/v_0)$ with $v_0\approx 4.4\kms$ (see Fig.~2). As discussed by Lefloch et al. (2012) and G\'omez-Ruiz et al. (2015), this slope is the specific signature of the outflow cavity associated with L1157-B1 ($g_2$). A blueshifted wing is detected up to $\approx -20\kms$ in the PN $J$=2--1 line profile (see Fig.~2), and we find a very good match with the SiO $J$=2--1 line profile obtained at a similar angular resolution of $27\arcsec$ with the IRAM 30m telescope. This suggests that the PN and SiO spatial distributions are very similar, and that PN comes from the region where the impact of the shock on the surrounding material is violent, with velocities larger than $\approx  25\kms$.
This provides us with an estimate for the size of the PN emitting region $\approx 18\arcsec$.

\begin{figure}
\includegraphics[width=0.9\columnwidth]{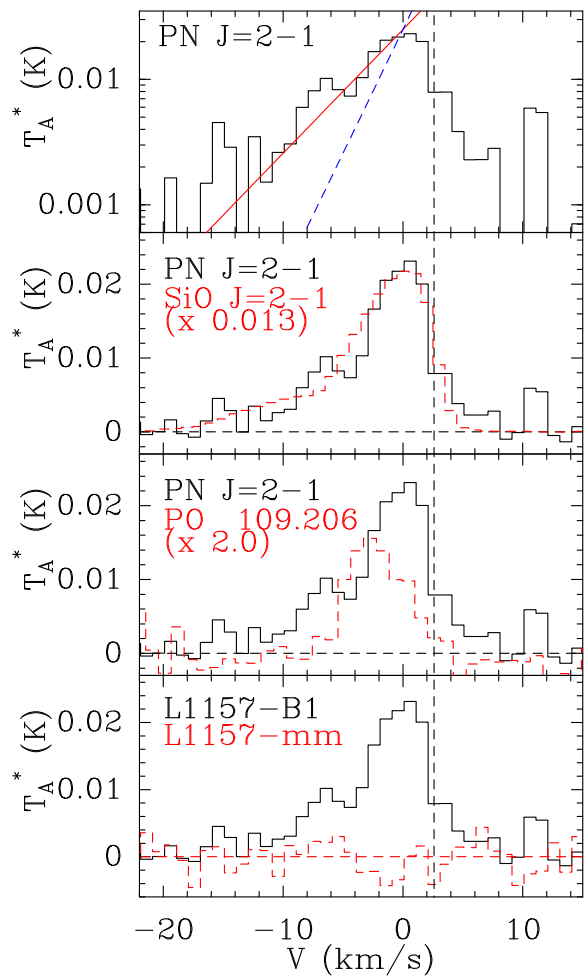}
\caption[]{From top to bottom: (a) Spectral profile of the PN $J$=2--1 displayed on a lin-log scale. We have superposed in red (blue) a fit to the spectral slope of the type $T_{\rm A}^{*}(v)\propto$ exp(v/v$_0$) with $v_0$= 4.4$\kms$ (2.2$\kms$), corresponding to the signature of components $g_2$ ($g_3$) of the outflow. (b) Superposition of the PN $J$=2--1 (solid) and SiO $J$=2--1 (dashed red) line profiles. A scaling factor of 0.013 was applied to SiO $J$=2--1 so to match the PN $J$=2--1 emission peak. (c) Comparison of the PN $J$=2--1 (solid) and PO 109.206 GHz (dashed red) lines. (d) Comparison of the PN $J$=2--1 line emission towards L1157-B1 (solid) and L1157-mm (dashed red). The vertical line in dashed marks the ambient cloud velocity $v_{lsr}$= $+2.6\kms$.}
\end{figure}

\subsection{PO}
We searched for all PO transitions in the ASAI bands. We used the spectroscopic parameters for phosphorus monoxide provided by the CDMS database (Kawaguchi et al. 1983, Bailleux et al. 2002).
A grand total of twelve PO transitions have  $A_{ij} \geq 10^{-5}\smu$ between 80 and 272 GHz. All the transitions fulfilling this criterion in the 3mm and 1.3mm bands were detected. This is the first detection of PO in a solar-type star forming region. Overall, lines are rather weak, with typical intensities of 5--8mK.  In the 2mm bands, two  line doublets have $A_{ij}$ larger than $10^{-5}\smu$ . Unfortunately,
 the higher rms noise in that part of the spectrum (5 mK per interval of $1\kms$) permits marginal detection of one doublet only. As a result, the flux of the 2mm transitions is rather uncertain, and we did not take them into account into the subsequent analysis to determine the excitation conditions of the PO gas.
The large number of transitions detected (10) makes us confident in the identification of PO in the line spectrum of L1157-B1.
The PO line emission peak is located close to $\approx -2\kms$, i.e. it is shifted from PN by $\sim 2$--$3\kms$ (see Fig.~2).  The velocity difference between the PO and PN  emission peaks implies that both species probably form in different regions of the shock. We note that the full width at zero intensity (FWZI)  is approximately $10\kms$: it is narrower for PO than for PN ($20\kms$).

\begin{table*}
{\small
%\tabletypesize{\scriptsize}
%\rotate
\caption{Spectroscopic and observational parameters of the molecular transitions of PO and PN observed towards L1157-B1, whose Einstein coefficient of spontaneous emission A$_{ij}$ is larger than $10^{-5}\smu$. Uncertainties on the line parameters are given in brackets. Line intensity uncertainties  are measured in a velocity interval of $1\kms$). Upper limits are estimated for  a $3\sigma$ rms.}
\label{tab:PN}
\begin{tabular}{lcclccrrrrrr}
\hline
\multicolumn{3}{c}{Transition}   & Frequency &$E_{up}$&HPBW   &F$_{eff}$ & B$_{eff}$& $\int T_{A}^{*}dv$ & $V$     & $\Delta V$ & $T_{A}^{*}$ \\
                   &     &        & (MHz)   &         (K)  & ($\arcsec$) &          &          &  (mK $\kms$)       & ($\kms$)& ($\kms$)   & (mK)   \\
\hline
\multicolumn{3}{l}{PN} &  &  &   &  &  &   & &  \\
\multicolumn{3}{l}{2 -- 1} & 93979.7689 & 6.8  & 26.2 & 0.95 & 0.80  & 148(13)& --0.24(0.25)& 5.97(.70) & 23.3(3.0) \\
\multicolumn{3}{l}{3 -- 2}  &140967.6921 & 13.5 & 17.5 & 0.93 & 0.74  & 186(13)& --0.67(0.23)& 6.27(.54) & 27.8(3.4)  \\
\multicolumn{3}{l}{5 -- 4 } &234935.2619 &  33.8 & 10.5 & 0.91 & 0.58  & 43(9)  & --0.62(.48) & 4.31(.88) & 9.4(1.5) \\
\multicolumn{3}{l}{6 -- 5}  &281914.2033 &  47.4 & 8.7  & 0.88 & 0.49  & 28(9)  & --1.47(.97) & 4.61(2.05)& 5.6(2.5)  \\ \hline
\multicolumn{3}{l}{PO   ${}^2\Pi_{1/2}$ }&      &      &       &  &  &  &  &   \\
J -- (J -- 1)    & Parity  & F--(F-1)&                    &         &        &         &         &               &                 &                 &      \\
\cline{1-3}    &                    &         &        &         &         &               &                 &                 &     \\
5/2 -- 3/2      &  e        & 3 --2     & 108998.445  &  8.4  & 22.6 & 0.95 & 0.79  & 63.2(7.0)&--2.18(.37) & 7.38(.94)  & 8.0(1.1)  \\
5/2 -- 3/2      &  e        & 2 -- 1      &109045.040  &  8.4  & 22.6 & 0.95 & 0.79  & 38.5(7.0)&--3.19(.73) & 7.80(1.60) & 4.6(1.2)  \\
5/2 -- 3/2      &  f         & 3 -- 2     &109206.200  &  8.4  & 22.5 & 0.95 & 0.79  & 27.3(6.0)&--2.34(.37) & 5.63(.84)  & 7.9(1.3)  \\
5/2 -- 3/2      & f          & 2 -- 1     &109281.189  &  8.4  & 22.5 & 0.95 & 0.79  & 24.5(7.0)&  0.20(.63) & 5.18(1.90) & 4.5(1.1)  \\
%$7/2_{-1,3,1/2}$ -- $5/2_{1,3,1/2}$      &152389.090  &  15.7 & 16.1 & 0.93 & 0.72  & -        & -          & -          & -   & \\
7/2 -- 5/2     & f          & 4 -- 3      &152656.979  &  15.7 & 16.1 & 0.93 & 0.72  & 118.0(19) & --1.34(.26) & 5.60(.99)  & 19.8(5.4)  \\
7/2 -- 5/2     & f          & 3 -- 2      &152680.282  &  15.7 & 16.1 & 0.93 & 0.72  &  66.6(17)  & --1.95(1.04) & 5.60 (.00)  & 11.2(5.0) \\
7/2 -- 5/2     & e          & 4 -- 3     &152855.454  &  15.7 & 16.1 & 0.93 & 0.72  &  <35 &  --               &  --      &  <15.0 \\
7/2 -- 5/2     & e          & 3 -- 2      &152888.128  &  15.7 & 16.1 & 0.93 & 0.72  &  <35  &  --                &   --     & <15.0  \\
%$7/2_{1,3,1/2}$ -- $5/2_{-1,3,1/2}$      &152953.253  &  15.7 & 16.1 & 0.93 & 0.72  & -        & -          & -          & -   & \\
%$11/2_{-1,5,1/2}$ -- $9/2_{1,5,1/2}$     &239704.364  &  36.7 & 10.3 & 0.91 & 0.58  & -        & -          & -          & -   & \\
11/2 -- 9/2   & f           & 6 -- 5     &239948.978  &  36.7 & 10.3 & 0.91 & 0.58  & 51.1(6.0)& --1.54(.30)& 5.06(.65) & 10.0(1.9) \\
11/2 -- 9/2   & f          & 5 -- 4      &239958.096  &  36.7 & 10.3 & 0.91 & 0.58  & 47.1(6.5)& --1.45(.35)& 4.86(.79) & 9.1(1.9) \\
11/2 -- 9/2   & e          & 6 -- 5     &240141.054  &  36.7 & 10.2 & 0.91 & 0.58  & 14.5(5.3)& --3.04(.43) & 2.25(.98) & 6.2(2.2)  \\
11/2 -- 9/2   & e          & 5 -- 4     &240152.530  &  36.7 & 10.2 & 0.91 & 0.58  & 28.5(9.0)& --4.89(.85) & 4.42(1.90)&6.1(2.2)  \\
%$11/2_{1,5,1/2}$ -- $9/2_{-1,5,1/2}$     &240268.285  &  36.7 & 10.2 & 0.91 & 0.58  & -        & -          & -          & -   & \\
\hline
\multicolumn{3}{l}{PH$_3$}             &                    &           &       &        &       &  &  &  &   \\
\multicolumn{3}{l}{$1_0$ -- $0_0$ } & 266944.662  &  12.8 & 9.2 & 0.89 & 0.52 &  <32 &  -         & -          &  <9.6(*) \\
\hline
\end{tabular}
(*) per interval of 2 MHz ($2.25\kms$)
}
\end{table*}

\subsection{PH$_3$}
We used the spectroscopic parameters for PH$_3$  provided by the CDMS database (M\"uller et al. 2013).
All the rotational transitions of PH$_3$ accessible in the ASAI bands are characterized by very low Einstein coefficients $A_{ij}$, typically less than $10^{-7}\smu$ and very high upper level energies $E_{up}$ (typically higher than $60\K$). The {\em only} transition with favourable excitation conditions
is the  $1_0$--$0_0$ at 266.944662 GHz, which has an $E_{up}$= $12.8\K$ and an $A_{ij}$ of $3.808\times 10^{-4}\smu$. We failed to detect this transition down to an rms of 3.2 mK ($\rm T_{A}^{*}$) in an interval of 2 MHz. Adopting a typical linewidth of $5\kms$, we obtain a $3\sigma$ upper limit of 32 mK$\kms$ (in $T_{A}^{*}$) for the line flux.

A search for other P-bearing molecules yielded only negative results.
%\section{Discussion}

\section{Molecular abundances}
%Our spectral line analysis suggests that the PN emission traces the same region as SiO $J$=2--1 in the outflow cavity of L1157-B1.
Plateau de Bure observations by Gueth et al. (1998) indicate a typical, round size of $18\arcsec$ for the SiO $J$=2--1 line emission, centered close to the nominal position of L1157-B1.
%interferometric observations are needed to determine whether the emission arise from the whole outflow cavity walls or from a reduced, compact %region.
Observations with the IRAM Plateau de bure interferometer at $3\arcsec$ resolution reveal an inhomogeneous structure with the presence of three  compact clumps of gas in L1157-B1, named "B1a-b-c", with a typical size of $4\arcsec$ (Benedettini et al. 2013).
We have considered two cases in the derivation of the physical conditions and abundance of PO and PN: a) Phosphorus emission arises from the whole shock region detected in SiO $J$=2--1 (size= $18\arcsec$), as suggested in Sect.3.2; b) Phosphorus emission arises from one of the compact clumps present in the SiO shock region (size= $4\arcsec$). We adopted a linewidth $\Delta V$= $6\kms$, in good agreement with the value obtained from a gaussian fit to the PO and PN line profiles of the $J$=2--1 and 3--2 transitions.
The corresponding fluxes are listed in Table~1.

\subsection{PN}
We derived the physical conditions in the PN gas from an analysis in the Large Velocity Gradient (LVG) approximation with the radiative transfer code MADEX. We used the PN-He collisional coefficients from Tobola et al. (2007) and  scaled by a factor of 1.37 to take
into account the difference in mass of H$_2$.

We first consider the case a) for the origin of PN emission.  We find as best fit solution a gas temperature $T_{\rm kin}$= $60\K$, n($\htwo$)= $9\times 10^4\cmmt$, and a source-averaged column density N(PN)= $9.0\times 10^{11}\cmmd$. The fit of the PN transitions is superposed in red on the line spectra in Fig. 1. A minimum $\chi^2$ analysis shows that best fitting solutions are obtained for N(PN)= $(9.0\pm 1.0)\times 10^{11}\cmmd$; the kinetic temperature is not well constrained and solutions in the range 40--$80\K$ are possible. On the contrary, the density is well constrained in the range (0.5-1.0)$\times 10^5\cmmt$. We note that these physical conditions are consistent with the gas kinetic temperature and the density previously determined in the L1157-B1 cavity, from a CO and CS multi-transition analysis (Lefloch et al. 2012; G\'omez-Ruiz et al. 2015).
Lefloch et al. (2012) estimated a CO gas column density of $1.0\times 10^{17}\cmmd$ for the L1157-B1 outflow cavity. Adopting a standard abundance ratio [CO]/[$\htwo$]= $10^{-4}$, we derive the abundance of PN relative to \htwo\ [PN]$\simeq 0.9\times 10^{-9}$. This value is in reasonable agreement with the previous determination of Yamaguchi et al. (2011):(0.2-0.6)$\times 10^{-9}$.

We now consider the case b), i.e. the alternative situation where PN arises mainly from one of the shocked gas clumps reported by Benedettini et al. (2013). We then obtain solutions of the type N(PN)= $4\times 10^{13}\cmmd$, $n(\htwo)$= $1\times 10^4\cmmt$, T=55K. This would result in a typical abundance [PN]$\simeq 4\times 10^{-8}$. Such low densities are hard to reconcile with the overall density  structure of L1157-B1, as derived by G\'omez-Ruiz et al. (2015), or even the density estimates obtained towards the compact clumps reported by Benedettini et al. (2013).
Our analysis favors case a):  PN arises from an extended region in the shock, close to the apex of the outflow cavity, where efficient grain sputtering is taking place.

\subsection{PO}
Due to the absence of coefficients available for PO-\htwo\ collisions, we carried out a simple LTE analysis using CASSIS.
The column density and the excitation temperature were determined from a best fit analysis ($\chi^2$) on the 5/2--3/2 transitions at 3mm and 1.3mm.  We could check a posteriori that the lines are optically thin.

Proceeding as above, we first assumed a size of $18\arcsec$ for the PO emitting region. The best fit solution is obtained for a rotational temperature $T_{\rm rot}$= $12.0\pm 0.9 \K$ and a source-averaged column density N(PO)= $(2.3\pm 0.4)\times 10^{12}\cmmd$, resulting into a gas phase abundance [PO]$\simeq$  $2.5\times 10^{-9}$. This is about 3 times  the abundance of PN. The fit to  the PO transitions  is superposed in red on the line spectra in Fig.~1, with a fixed Full Width at Half Maximum (FWHM) of $5.5\kms$   and $v_{lsr}$ of $-2.5\kms$, with the above parameters.

Like for PN, we consider the case where the PO emitting region would arise from a compact shock region. Not surprisingly, a larger column density is required to account for the PO emission: typically $3\times 10^{13}$ and $1.3\times 10^{14}\cmmd$ for a source size of $4\arcsec$ and $2\arcsec$, respectively. This corresponds to a PO abundance [PO]= $3\times 10^{-8}$ and $1.3\times 10^{-7}$, respectively.

\subsection{PH$_3$}
We used CASSIS to obtain an upper limit on the abundance of phosphine from the non-detection of the ground transition $1_0$--$0_0$ at 266944.662 GHz.
We adopted a typical excitation temperature of $10\K$, a linewidth of $5\kms$, and adopted a typical source size of $18\arcsec$.  Taking into account the rms of 3mK, comparison with the ASAI data shows that N(PH$_3$) has to be less than $\approx 10^{12}\cmmd$, which results into an upper limit of $10^{-9}$ for the phosphine abundance in the gas phase.

As a conclusion, a careful determination of the abundance of P-bearing species in the gas phase shows that they contain {\em only} a few percent of the total elemental phosphorus.

\section{Phosphorus chemistry in the L1157-B1 shock}
In Section 3, we showed that the emission from PN and PO arises from the L1157-B1 outflow cavity, which can be described by a C-type shock (Lefloch et al. 2012; Holdship et al. 2016). We therefore investigate the origin of the PO and PN emission in this region by the use of the chemical gas-grain model UCL\_CHEM (Viti et al. 2004a) coupled with the parametric C-shock model by Jim\'enez-Serra et al. (2008). We use the same approach as Viti et al. (2011) where we modelled the ammonia and water emission across the same region.
We note  that a similar approach was used by Aota \& Aikawa (2012) to reproduce the previous observations of PN and PO performed by Yamaguchi et al. (2011) toward the same outflow cavity. We stress that our new observational constraints lead to different conclusions about the history of the region and the shock parameters. This point is more specifically addressed in Sect.~5.3.

%of L1157 B1 by using a chemical model coupled with the same shock module we use here. As we shall see, we are in %agreement with their findings; however Aota \& Aikawa (2012) were looking for a solution whereby PN was %abundant, while PO was not, as prior to the present observations Yamaguchi et al. (2011) detected PN in L1157-B1 %but failed to detect PO. CONNECT THIS TO PREVIOUS SECTIONS.
We study the evolution of phosphorous-bearing species in a one dimensional C-shock, but with the specific aim of determining whether PO and PN are mainly formed and destroyed in the gas phase prior to the shock passage, on the grains, or during the shock event(s). Moreover, as in Viti et al. (2011), our model is able to determine whether the differences in line profiles discussed above may be explained by the differences in abundances at different velocities.

\subsection{Modelling}
A detailed description of UCL\_CHEM coupled with the shock model can be found in Viti et al. (2011). Here we briefly summarize the main characteristics of the code. The code is run in two phases, where Phase I forms a dense core out of a diffuse medium, starting from an essentially atomic gas. We adopt an initial density of 100 cm$^{-3}$ for the diffuse medium. During this phase, gas-phase chemistry, freezing on to dust particles and subsequent surface processing occur. The sticking efficiency for all species is assumed to be 100\% but the rate of depletion is a function of density (in a similar manner as in Rawlings et al. 1992). The density at the end of Phase I is a free parameter, called from now on the pre-shock density.
The second phase computes the time-dependent chemical evolution of the gas and dust during the passage of a shock.
The model includes both thermal desorption, due to the dust being heated by the presence of the outflow, as well as sputtering of the icy mantles once the dynamical age across the C-shock has reached the "saturation timescales", as in Jimenez-Serra et al. (2008). The saturation timescale corresponds to the time when most of the ices have been injected into the gas phase, due to the passage of the shock (Jimenez-Serra et al. 2008). Such  timescales are inversely proportional to the density of the gas and for pre-shock \htwo\ densities of $10^4$--$10^5\cmmt$, these times range between 10 and 100 years, i.e. factors 10-100 shorter than the typical dynamical ages of young molecular outflows ($10^3\yr$).
Our chemical gas-phase network is taken from UMIST 12\footnote{http://udfa.ajmarkwick.net/}and is augmented with updates from the KIDA database\footnote{http://kida.obs.u-bordeaux1.fr/}. The surface network comprises mainly hydrogenation reactions. In particular we note that neither UMIST 12 nor KIDA contains a network for PH$_3$. Rather than creating an {\it ad hoc} set of reactions for this species we made the assumptions that this species will be preferentially formed on the grains by hydrogenation and that therefore PH$_2$ could act as a proxy for PH$_3$. Of course this also means that we are assuming that the routes for gas phase destruction for PH$_3$ and PH$_2$ are similar. In all our figures we therefore present PH$_2$ as a proxy for PH$_3$.

We ran a small grid of models where we varied:
\begin{itemize}
\item the pre-shock density; we have adopted two possible values for the total hydrogen nuclei density n(H+$2\times$ H$_2$)based on previous studies of this cavity: 10$^4$ and 10$^5\cmmt$.
\item the shock velocity, from 20 to $40\kms$. From the pre-shock density and the shock velocity the maximum temperature of the neutral fluid attained within the shock is taken from Figures 8b and 9b in Draine et al. (1983).
\item the initial elemental abundance of phosphorous: while its abundance in the diffuse, warm ISM, is found to be solar (e.g. Jenkins, Savage \& Spitzer 1986), all the studies so far of phosphorous-bearing species in star forming regions  have found that the initial elemental abundance of phosphorous needs to be depleted by up to a factor of 100 in order to match the observations (Aota \& Aikawa 2012; Fontani et al. 2016; Rivilla et al. 2016). We have therefore varied this parameter and adopted values of solar (2.57$\times$10$^{-7}$) down to a depletion factor of 100.
\item the duration of the pre-shock phase (Phase I). We have considered a short-lived and a long-lived scenario, respectively, to investigate the impact of different initial gas compositions, as a result of different cloud ages. In the first case, the core  is immediately subjected to the passage of a shock when it reaches the pre-shock density; in the second case, the core is allowed to remain static for about 2 million years.
\end{itemize}

\subsection{Results}

\begin{figure*}
\includegraphics[width=1.8\columnwidth]{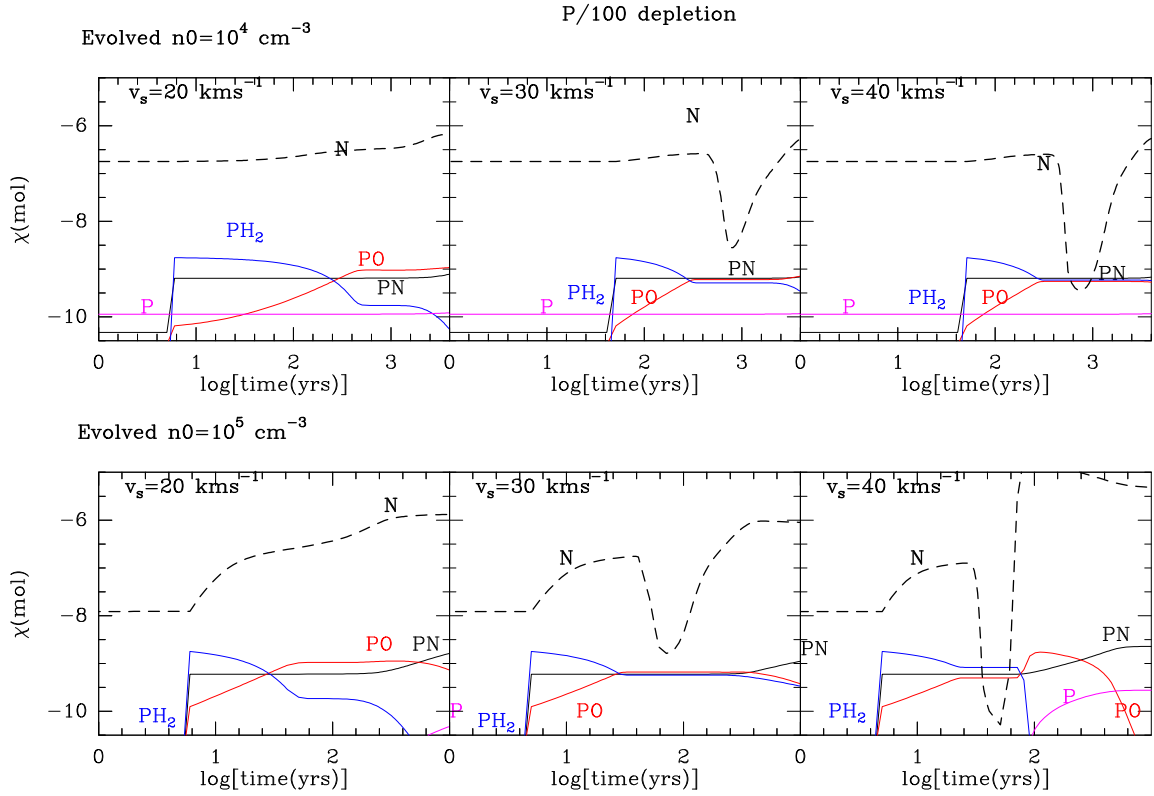}
\caption[]{Fractional abundances of selected species as a function of time for the shocked phase. In all models the initial elemental abundance of phosphorous is depleted by a factor of 100 with respect to its solar abundance. The pre-shock phase is long-lived.}
\end{figure*}

\begin{figure*}
\includegraphics[width=1.8\columnwidth]{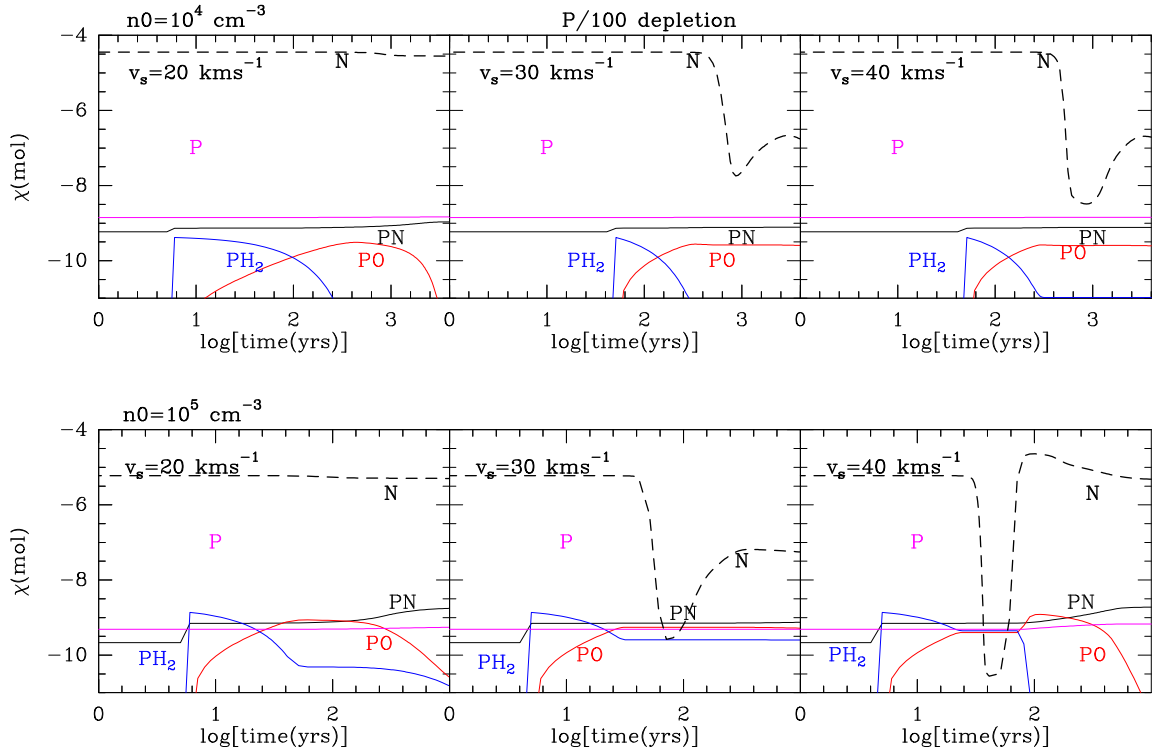}
\caption[]{Fractional abundances of selected species as a function of time for the shocked phase. In all models the initial elemental abundance of phosphorous is depleted by a factor of 100 with respect to its solar abundance.
The pre-shock phase is short-lived.}
\end{figure*}

The observations yield abundance values for PO and PN of the order of 1--3$\times 10^{-9}$, for a common source size of $18\arcsec $, with PO larger by a factor of 3 or so than PN. PH$_3$ is undetected, and has an upper limit of 10$^{-9}$.  We present in Fig.~3 the evolution of phosphorous bearing species as well as nitrogen as a function of time during the passage of the shock (Phase II), when the initial elemental abundance of phosphorus is depleted by a factor of 100 and the pre-shock phase is long-lived.  We show the result for three  shock velocities (20,30, $40\kms$), pre-shock  densities of $10^4$ and $10^5\cmmt$, and long-/short-lived pre-shock phase in Figs.~3 and 4, respectively. We note that the magnitude of the depletion effect affects only the overall abundance of P-bearing species.
%The other cases (short-lived pre-shock phase and P depletion factor of 100,  no P depletion and short/long-lived pre-shock phase)  are %presented in Figs.~A1-2-3.

Our first result is that none of the models where phosphorous is solar can fit the observations, as all phosphorous bearing species are about two orders of magnitude higher than what is observed.  Models where the total gas-phase phosphorous is $\sim 10^{-9}$ are therefore favoured.

As can be seen in Fig.~3, we find that PH$_2$ is immediately released in the gas phase, as a result of the grain mantle sputtering at the passage of shock. PN follows closely the evolution of  PH$_2$, while PO forms gradually and its abundance rises slowly  in the shock. This is in agreement with our observations, where there is a clear shift toward more blue-shifted velocities for PO than for PN, which is consistent with a delay in the formation of PO in the gas phase with respect to PN.

If the pre-shock phase (Phase I) is short-lived (Fig.~4) we never reach a situation where PO is clearly larger than PN throughout the shock evolution.  As explained below, this is due to the lack of significant depletion of atomic N for time-scales <1 Myr. Atomic phosphorus also remains abundant in the gas phase, at abundances of about $10^{-9}$, comparable to or larger than that of PN, depending on the pre-shock density.
If the pre-shock phase (Phase I) is short-lived we never reach a situation where PO is clearly larger than PN throughout the shock evolution. This is the reason why Aota \& Aikawa (2012) opted for a short-lived pre-shock phase to explain the upper limits of PO measured toward L1157-B1 by Yamaguchi et al. (2011).
Given the uncertainties in the derivation of PO and PN abundances, we cannot discard a situation the PO/PN ratio is simply $\approx 1$, however.

A chemical analysis on the formation and destruction of PO and PN shows that the key player in their chemistry is atomic nitrogen (as also found by Aota \& Aikawa 2012), which in turn is mainly released from ammonia NH$_3$ in the shock: for example, for the higher pre-shock densities ($10^5\cmmt$) and weak shocks ($20\kms$), a significant fraction of atomic nitrogen is still present in the gas phase and hence PO is destroyed by the reaction  $$ N + PO \rightarrow PN $$.

This reaction contributes to $\sim$85\% to the total destruction of PO.
When the shock velocity is higher, about $30 \kms$, the temperature of the neutral fluid is also higher and favours the conversion of N into NH$_3$, and therefore PO survives in the gas phase for longer.  However, at higher velocities ($40\kms$), as also shown in Viti et al. (2011), NH$_3$ is efficiently destroyed in the post-shock gas, liberating N into the gas phase and augmenting the destruction of PO via N (a route that here contributes to 100\% of the PO destruction). This leads to a faster drop for PO. This process requires that the maximum temperature in the shock is larger than $4000\K$.

Because the amount of available nitrogen is key to the survival of PO, a higher pre-shock density favours it: for higher densities, more nitrogen is locked onto grains in the form of NH$_3$.  The atomic Nitrogen abundance decreases by $\approx 10$ when the pre-shock density increases from $10^4$ to $10^5\cmmt$ (see e.g. Figs.~3--4).
 Since from previous discussion in Sect.~3.2,  the terminal velocity of PO is smaller (about $-10\kms$) than the PN one (almost $-20\kms$), the model with a shock velocity of $40\kms$ (bottom right of Fig.~3) is the best match, as PO is destroyed at the end of the post-shock region as a consequence of the destruction of NH$_3$.

The best matching models are in fact those where the pre-shock phase is longer-lived (Fig.~3), again, because atomic nitrogen has more time to get locked in the icy mantles before the arrival of the shock. In all the models from Fig.~3, PN and PO are comparable for most of the dissipation length.
%, with, again, the $40\kms$ shock with a pre-shock density  of $10^5\cmmt$  being the best match if one takes into consideration the respective terminal velocities. %A longer lived pre-shock clump was also favoured by the models of Aota \& Aikawa (2012).
The only model for which we find a sharp decrease in abundance in the postshock region is the model with pre-shock density of $10^5\cmmt$ and a shock velocity of $40\kms$.  Such a drop occurs at a time of $\sim 600\yr$ and translates into lower terminal velocities of the PO line emission. Models with a shock velocity of $20\kms$ and $30\kms$  also indicate a PO abundance drop from the gas phase. This abundance drop occurs  at a longer time however, once passed the dissipation length, which makes it difficult to quantify accurately based on our simple modelling. Because of this, and because similar shock velocities are also required to explain the different line profiles observed toward L1157-B1 for NH$_3$ and H$_2$O (Viti et al. 2011), we favour the scenario in which PN and PO are formed in a shock with a velocity of $40\kms$.
Finally, we note that a high pre-shock density also favours the idea that dense clumps along outflows pre-exist the outflow event itself and are not in fact formed by compression due to the arrival of the shock (Viti et al. 2004b).

\subsection{Comparison with previous results}
Using the detection of the PN $J$=2--1 line and the non-detection of PO by Yamaguchi et al. (2011) in L1157-B1, Aota \& Aikawa (2012) modelled the phosphorous chemistry in the shocked region of L1157-B1 by using a chemical model coupled with the parametric shock model  by Jimenez-Serra et al. (2008). These authors searched for shock solutions whereby PN was abundant while PO was not. The wider and richer spectral content of the ASAI  data brings more robust constraints on the properties of P-bearing species PO, PN (and PH$_3$) in L1157-B1, affecting some of  their conclusions on the shock properties: their study favours a shock velocity of $20\kms$, while our results suggest that PN and PO are formed in a $40\kms$ shock. In our model a maximum temperature of 4000 K is indeed required in the shock to liberate atomic N into the gas phase from NH3 and to destroy PO at higher velocities in the shock (as suggested by the observed different terminal velocities of PN and PO; Section 3.2). Our conclusions on the depletion of elemental phosphorus by a factor of 100 in molecular dark clouds, and of the importance of atomic N in the chemistry of PO and PN, agree with those of Aota \& Aikawa (2012).

Recent studies of PN and PO towards the high-mass star forming regions W51 e1/e2 and W3(OH) by Fontani et al. (2016) and Rivilla et al. (2016) reach conclusions similar to ours: phosphorus appears to be depleted in quiescent molecular gas by more than one order of magnitude.  The abundances they derive towards these massive sources are of the order $10^{-10}$, typically one order of magnitude less than towards the shock L1157-B1. Interestingly, they report similar PO/PN abundance ratios ($\simeq$ 2--3).  Observations with the NOEMA array would permit more accurate determination of the depletion factor by resolving the size of the emitting region of PO and PN at a few arcsec scale.
Mapping the relative spatial distribution of PO and PN as a function of velocity would permit confirmation of our conclusions on the shock parameters.

\section{Conclusions}
We report on a systematic search for P-bearing species towards the solar-type star forming region L1157. We have unambiguously detected   emission from PN and, for the first time, from the pre-biotic molecule PO in the outflow shock region L1157-B1.  No emission from P-bearing species was detected towards the envelope of the protostar L1157-mm. Spectral line profile analysis suggests that the emission originates from the same region as SiO, with a typical size of 15--$20\arcsec$. The abundances of PO and PN  are found comparable, with values of 2.5 and $0.9\times 10^{-9}$, respectively.

A simple modelling using the C-shock code of Jimenez-Serra et al. (2006) coupled  the chemical gas-grain model UCL\_CHEM (Viti et al. 2004) allows us to reproduce the main features of PO and PN emission in the shock. Our main conclusions are as follows:\\
- Phosphorus is depleted by about a factor 100 in the gas phase. \\
- Atomic Nitrogen plays a key role in the formation and destruction routes of PO and PN\\
- The observed PO/PN abundance ratio is $\approx 3$, and brings constraints on the duration of the pre-shock phase, which has to be larger than $\simeq 10^6\yr$ and the pre-shock density, of the order of $10^5\cmmt$. \\
- The maximum temperature in the shock has to be  $\sim 4000\K$ in order to account for the difference of terminal velocities between PO and PN. \\

Follow-up observations at arcsec scale with the NOEMA array would permit probing the model we propose for the formation of PO and PN in the shock. By resolving the size of the emitting region of PO and PN,  it would possible to confirm that both species form in different regions of the shock, and to estimate more accurately the  Phosphorus depletion factor in the gas phase.

\section*{Acknowledgements}
Based on observations carried out as part of  the Large Program ASAI (project number 012-12) with the IRAM 30m telescope.
IRAM is supported by INSU/CNRS (France), MPG (Germany) and IGN (Spain).  This work  was supported by the CNRS program "Physique et Chimie du Milieu Interstellaire" (PCMI) and by a grant from LabeX Osug@2020 (Investissements d'avenir - ANR10LABX56). E. M. acknowledges support from the Brazilian agency FAPESP (grant 2014/22095-6 and 2015/22254-0). I.J.-S. acknowledges the financial support received from the STFC through an Ernest Rutherford Fellowship (proposal number ST/L004801/1).

%%%%%%%%%%%%%%%%%%%%%%%%%%%%%%%%%%%%%%%%%%%%%%%%%%



%\bibliographystyle{mnras}

\begin{thebibliography}{99}
%\bibitem[]{} Arce, H., et al., 2008, ApJ,
\bibitem[]{} Altwegg, K., Balsiger, H., Bar-Nun, A., et al., 2016, Science Advances, 27 May 2016, 2, 160085, 
\bibitem[]{} Aota,T., Aikawa, Y., 2012, ApJ, 761, 74
\bibitem[]{} Agundez, M., Cernicharo, J., Gu\'elin, M., 2007, ApJ, 662, L91
\bibitem[]{} Asplund, M., Grevesse,N., Sauval, J., Scott, P., 2009, ARA\&A, 47, 481
\bibitem[]{} Ceccarelli, C., Bacmann, A., Boogert, A., et al., 2010, A\&A, 521, L22
\bibitem[]{} Benedettini, M., Viti, S., Codella, C., et al., 2013, MNRAS, 436, 179
\bibitem[]{} De Beck, E., Kam\i\'{n}ski, T., Patel, N.A., et al., 2013, A\&A, 558, 132
\bibitem[]{} Draine,B.T., Roberge, W.G., Dalgarno, A., 1983, ApJ, 264, 485
\bibitem[]{} Dufton, P.L., Keenan F.P., Hibbert, A., 1986, A\&A, 164, 179
\bibitem[]{} G\'omez-Ruiz, A., Codella, C., Lefloch, B., et al., 2015, MNRAS, 446, 3346
\bibitem[]{} Fontani, F., Rivilla, V.M., Caselli, P., et al., 2016, ApJ, 822, L30
\bibitem[]{} Gueth, F., Guilloteau, S., Bachiller, R., 1998, A\&A, 333, 287
\bibitem[]{} Halfen,D.T., Ilyushin, V., Ziurys, L.M., 2011, ApJ, 743, 60
\bibitem[]{} Jenkins, E.B., Savage, B.D. , Spitzer, L., 1986, ApJ, 301, 355
\bibitem[]{} Jimenez-Serra, I., Caselli, P., Martin-Pintado, J., Hartquist, T.W., 2008, A\&A, 482, 549
\bibitem[]{} Lefloch, B., Cabrit, S., Busquet, G., et al., 2012, ApJ, 757, L25
\bibitem[]{} Macia, E., 2005, Chem Soc. Rev., 34, 691
\bibitem[]{} Mendoza, E., Lefloch, B., Lopez-Sepulcre, A., et al., 2014, MNRAS, 445, 1151
\bibitem[]{} Matthews, H.E., Feldman, P.A., Bernath, P.F., 1987, ApJ, 312, 358
\bibitem[]{} Rawlings, J.M.C., Hartquist, T.W., Menten, K.M., Williams, D.A., 1992, MNRAS, 225, 471
\bibitem[]{} Rivilla, V.M., Fontani, F., Beltran, M.T., et al., 2016, astroph:arXiv:1605.06109v2
%\bibitem[]{} Saladino, R., Botta, G., Pino, S., 2012, Chem. Soc. Rev., 41, 5526–5565
\bibitem[]{} Saladino, R., Botta, G., Pino, S., 2012, Chem. Soc. Rev., 41, 5526-5565
%\bibitem[]{} Sutton,E.C., Blake,G.A., Masson, C.R., Phillips,T.G., 1985, ApJS, 58, 341
\bibitem[]{} Tenenbaum, E.D., Woolf, N.J., Ziurys, L.M., 2007, 666, L29
\bibitem[]{} Tobola, R., Klos, J., Lique, F., et al., 2007, A\&A, 468, 1123
\bibitem[]{} Turner, B.E., Bally, J., 1987, ApJ, 321, L75
\bibitem[]{} Viti, S., Collings, M.P., Dever, J.W., 2004a, MNRAS, 354, 1141
\bibitem[]{} Viti, S., Codella, C., Benedettini, M., Bachiller, R., 2004, MNRAS, 350, 1029
\bibitem[]{} Viti, S., Jimenez-Serra, I., Yates, J.A., et al., 2011, ApJ, 740, L3
\bibitem[]{} Yamaguchi, T., Takano, S., Sakai, N., et al., 2011, PASJ, 63, L37
\bibitem[]{} Ziurys, L., 1987, ApJ, 321, L81
\end{thebibliography}


%%%%%%%%%%%%%%%%%%%%%%%%%%%%%%%%%%%%%%%%%%%%%%%%%%

%%%%%%%%%%%%%%%%% APPENDICES %%%%%%%%%%%%%%%%%%%%%

%\appendix

%%%%%%%%%%%%%%%%%%%%%%%%%%%%%%%%%%%%%%%%%%%%%%%%%%

% Don't change these lines
\bsp	% typesetting comment
\label{lastpage}
\end{document}